\documentstyle[psfig,times]{mn2e}

\title[Energetics - Spectral correlations vs the BATSE Gamma-Ray
  Bursts population] {Energetics - Spectral correlations vs the BATSE
  Gamma-Ray Bursts population} \author[Z. Bosnjak et al.]
  {\parbox[]{6.5in} {Z. Bosnjak$^1$, A. Celotti$^1$, F. Longo$^2$,
  G. Barbiellini$^2$}\\ $^1$S.I.S.S.A., via Beirut 2-4, I-34014
  Trieste, Italy\\ $^2$Department of Physics and INFN, via Valerio 2,
  I-34100 Trieste, Italy \\ }

\date{}

\begin{document}

\maketitle

\begin{abstract}
Recently proposed correlations between the energetics of Gamma-Ray
Bursts (GRB) and their spectral properties, namely the peak energy of
their prompt emission, are found to account for the observed fluence
distribution of all `bright' BATSE GRB. Furthermore for an intrinsic
GRB peak energy distribution extending toward lower energies with
respect to that characterizing bright GRB, such correlations allow to
reproduce the fluence distribution of the whole BATSE long GRB
population. We discuss the constraints that such analysis imposes on
the shape of the peak energy distribution, the opening angle
distribution and the tightness of such correlations.
\end{abstract}


\section{Introduction}
Among the most interesting clues on the physical processes taking
place in GRB are the recently proposed correlations between their
energetics and spectral properties.  More precisely it has been
suggested (Lloyd--Ronning, Petrosian \& Mallozzi 2000; Amati et al.
2002 [A02 hereafter]; Sakamoto et al. 2004; Lamb et al. 2004; Atteia
et al. 2004) that the apparent isotropic energy of the prompt
correlates with the intrinsic peak energy (in $\nu f(\nu)$) of the
integrated emission, with a dependence $E_{\rm iso}\propto E_{\rm
peak}^{0.5}$. A similar correlation has been found between $E_{\rm
peak}$ and the peak luminosity (Yonetoku et al.  2004).  More
recently, Ghirlanda, Ghisellini \& Lazzati (2004) (GGL04 hereafter),
by correcting for the putative fireball opening angle estimated from
the (achromatic) break time in the afterglow light curve (Sari, Piran
\& Halpern 1999; Frail et al. 2001; Bloom, Frail \& Kulkarni 2003)
argued that an even tighter correlation holds between the actual
energetic and $E_{\rm peak}$, namely $E_{\gamma}\propto E_{\rm
peak}^{0.7}$. Such correlations are based on (at most) the $\sim$ 40
long GRB for which redshift information is currently available.
Although no unique and robust interpretation of such results has been
found so far (e.g. Schaefer 2003; Liang, Dai \& Wu 2004; Eichler \&
Levinson 2004; Rees \& Meszaros 2005), it is clear that if these were
to hold for the whole GRB population (see Friedman \& Bloom 2004;
Nakar \& Piran 2004; Band \& Preece 2005 for dissenting views), they
could be powerful clues to the physical origin of the prompt emission
and have important repercussions on the cosmological use of GRB.

In order to test whether these relations characterize the bulk of the
GRB population (until a significantly larger number of redshift can be
determined), we tested their consistence against the observed peak
energy and fluence distributions, under the assumption that the GRB
events follow the cosmological star formation rate redshift
distribution. Although the found statistical consistency is not a
proof of such correlations, it supports the view that they are indeed
representing an intrinsic properties of all (BATSE) long GRB.

The outline of this Letter is the following: we detail our assumptions
and procedure in Section 2, present our results in Section 3 and
finally discuss them in Section 4.  Preliminary results of this work
have been presented by Bosnjak et al. (2004). While finishing writing
this Letter we received the manuscript from Ghirlanda, Ghisellini \&
Firmani (2005), who -- through a complementary and independent analysis
-- reach remarkably similar results to those reported in this work.

\section{Method and assumptions}

We aim at testing whether the observed peak energy and fluence
distributions are consistent with the A02 and GGL04 proposed
correlations.

More precisely, we considered the sample of BATSE GRB analyzed by
Preece et al. (2000) (referred to as the `bright' BATSE sample
hereafter), consisting of 156 events for which $E_{\rm peak}$ have
been estimated.  We then simulated -- via Monte Carlo method -- the
fluence distribution for a population of GRB characterized by the
corresponding $E_{\rm peak}$ distribution as follows:

\vskip 0.2 truecm
\noindent
-- assumed that the GRB population follows the star formation rate
distribution in redshift (as estimated by Madau \&
Pozzetti 2000), namely $R_{\rm GRB}(z) = 0.3 \exp(3.4 z) [\exp(3.8
  z)+45]^{-1}$ M$_{\odot}$ yr$^{-1}$ Mpc$^{-3}$;
\vskip 0.2 truecm
\noindent
-- adopted the observed bright BATSE GRB $E_{\rm peak}$ distribution,
as obtained by averaging the Preece et al. (2000) results of their
time resolved spectral analysis;
\vskip 0.2 truecm
\noindent
--randomly assigned a redshift and a characteristic $E_{\rm peak}$ to
each event;
\vskip 0.2 truecm
\noindent
--adopted the A02 correlation (and its spread) to estimate the
  corresponding energetics;
\vskip 0.2 truecm
\noindent
--by applying the cosmological corrections\footnote{Throughout this
  work we adopt a `concordance' cosmology $\Omega_{\Lambda}$ = 0.7,
  $\Omega_{\rm M}$ = 0.3, and $H_{0}$ = 65 km s$^{-1}$ Mpc$^{-1}$
  ($H_{0}$ = 70 km s$^{-1}$ Mpc$^{-1}$ for the GGL04 case).} estimated
  the corresponding fluence in the 50-300 keV energy range (a typical
  Band spectral representation with $\alpha=-1$ and $\beta =-2.25$ has
  been adopted, see Preece et al. 2000);
\vskip 0.2 truecm
\noindent
--compared the resulting fluence distribution with that of bright
BATSE GRB. The comparison of fluences clearly avoids the need of a
further assumption about the GRB durations.
\vskip 0.2 truecm

An analogous test by adopting the GGL04 relation is clearly less
straightforward, as it requires the information on the GRB opening
angle distribution.  The latter is however constrained only by 16 (8)
GRB for which an estimate (limit) on the opening angle can been
determined from the break time of the afterglow light curve (see
GGL04). We approximated such distribution as a lognormal function and
constrained it by requiring that it can reproduce the observed
fluences.

We then explored the possibility that the whole of the BATSE long GRB
population might follow such relations.  Clearly, if indeed these were
to hold, the adoption of the BATSE bright GRB peak energy distribution
biases the selection to typically high fluence events. In order to
account also for dim GRB we thus extrapolated the $E_{\rm peak}$
distribution toward lower energies.

The comparison between the simulated and observed fluences has been
performed by estimating the maximum difference $D$ in the cumulative
distributions, as in the Kolmogorov-Smirnoff (KS) test. The parameter
$D$ has been used to compare the agreement with data of the different
models (i.e. different assumptions/parameters), although formally the
associated probability of two distributions being drawn from the same
parent one would be only $P_{\rm KS}$= 0.002 for $D < 0.07$ (which we
treat as a limit for a qualitatively satisfactory agreement).

\section{Results}

The strongest finding is indeed that the simple adoption of the A02
correlation generates a fluence distribution consistent with the
observed one for bright BATSE GRB.  The comparison of the predicted
and observed distributions is shown in Figure~1 (top panel) and their
formal consistency is confirmed by a KS test (probability $P_{\rm
KS}$= 0.06).

The observed fluences can be satisfactorily reproduced ($P_{\rm KS}$=
0.18) also by adopting the GGL04 relation for an lognormal opening
angle distribution peaking around $\sim 4-5^o$ and mimicking the
distribution of the (few) estimated opening angles (see Figure~2).

\begin{figure}
\psfig{file=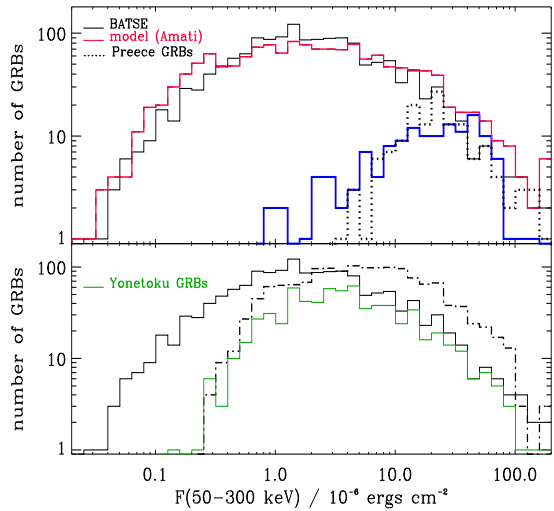,width=8.5truecm,height=7 truecm}
\caption{Fluence distributions for `bright' BATSE GRB (Preece et
al. 2000) (black dotted line), the whole of BATSE long GRB population
assuming: the A02 relation (top panel, blue and red lines); the GGL04
relation, assuming the `bright' GRB opening angle distribution for the
whole of the BATSE sample (black dot-dashed line, bottom panel); the
observed distribution of the sample by Yonetoku et al. (2004) (green
line, bottom panel).}
\end{figure}

In order to account for the fluences of the whole BATSE GRB population
\footnote{http://cossc.gsfc.nasa.gov/batse/BATSE$\_$Ctlg/flux.html}
($\sim 1500$ events), the intrinsic peak energy distribution has to
extend to lower energies, as the spread in fluences arising from the
cosmological distance of GRB would span a range much smaller than what
required. Furthermore, as the range in $E_{\rm peak}$ would dominate
in determining the range in fluences, the shape of the $E_{\rm peak}$
distribution should be qualitatively similar to that of fluences.

Indeed, in Figure~3 we report the $E_{\rm peak}$ distribution which
allows to satisfactorily reproduce the overall fluence distribution
(as shown in Figure~1, top panel) which broadly peaks around $\sim$ 80
keV.  Attempts have been made to determine how such extrapolation is
constrained, by considering also other $E_{\rm peak}$ distributions
(shown in Figure~3), namely a distribution increasing further and
peaking around $\sim 60$ keV and one peaking around $100-200$ keV, the
latter reproducing the distribution for the GRB examined by Yonetoku
et al. (2004).  Interestingly our analysis is quite sensitive to the
extrapolation, resulting in inconsistent fluence distributions for
either alternatives (by over and under estimating the dimmest GRB,
respectively (the result for the latter case is shown in Figure~1,
bottom panel).

However when the GGL04 correlation is adopted, the extrapolation to
lower $E_{\rm peak}$ shown in Figure~3 cannot account by itself for
the fluence distribution if the (narrow) distribution of angles
inferred for bright GRB is adopted.  Indeed, as shown in Figure~1
(bottom panel), the corresponding fluence distribution in such case
results to be a factor $\sim 5$ higher and narrower than the observed
one.  Within this scenario such discrepancy can be accounted for if
BATSE GRB include a large fraction of bursts with wider opening
angles: Figure~2 reports the inferred (lognormal) opening angle
distribution which yields a satisfactory agreement for the
fluences. This peaks around 6-8$^o$ and extends to about 20-25$^o$.
The larger central value of the angles has to be considered as a
representative parameter, which could in principle mimic other
effects, like possible absorption.

It should be stressed that both the extrapolated $E_{\rm peak}$ as
well as the opening angle distributions are quite constrained, both in
shape and in extent.

\begin{figure}
\psfig{file=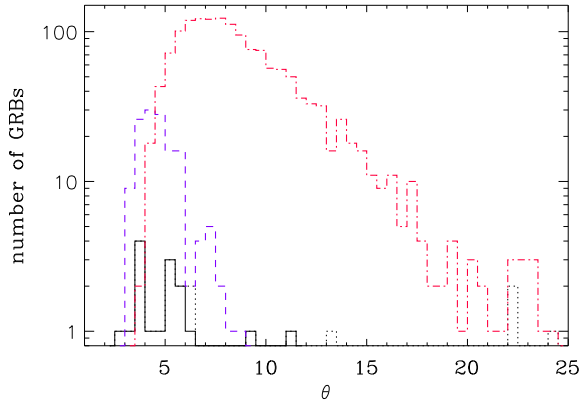,width=8.5truecm}
\caption{Opening angle distributions, as constrained by the request
that the GGL04 correlation is representative of bright BATSE GRB
(dashed line) and the whole of the BATSE GRB population
(dot-dashed). Reported are also the values inferred from the break
time of the afterglow light curves (solid histogram, data from
GGL04).}
\end{figure}

\begin{figure}
\psfig{file=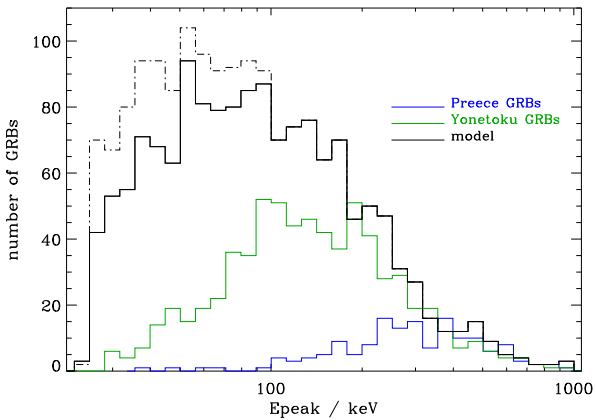,width=8.5truecm}
\caption{$E_{\rm peak}$ distributions for the bright BATSE GRB
(observed, Preece et al. 2000), for the GRB sample examined by
Yonetoku et al. (2004) and that adopted in this work for the whole
BATSE long GRB sample.  The dashed line shows the other $E_{\rm peak}$
distribution tested.}
\end{figure}

\subsection{Inferred properties of BATSE GRB}

Within the above assumptions, the population of GRB with properties
consistent with those of BATSE GRB can be characterized in terms of
redshift distribution and luminosity function.

Given that the redshift (up to $z\sim 5$) is not the primary driver
for the observed low fluence events, the sample basically follows the
assumed cosmological distribution. For this same reason, the analysis
did not give significantly different results (within a factor 2 in
fluences) when a star formation rate $\sim$ constant above $z\sim 2$
(case 2 in Porciani \& Madau 2001) was adopted.

The inferred `luminosity' function (in terms of $E_{\gamma}$),
reported in Figure~4, clearly reflects the $E_{\rm peak}$
distribution.  Interestingly, this well agrees with those deduced from
number counts constraints, providing a self-consistency check on the
assumptions imposed in the present analysis. In fact, the parameters
which characterize it (as lognormal function) are consistent with
those determined by Sethi \& Bhargavi (2001) and Schmidt (2001) in
terms of peak and width (including a decline at low $E_{\gamma}$) for
an average GRB duration of $\sim 100$ sec, and qualitatively
consistent with the characterization reported by Guetta, Piran \&
Waxman (2004) for the higher energy part in terms of a broken
power--law (see Figure~4).

\begin{figure}
\psfig{file=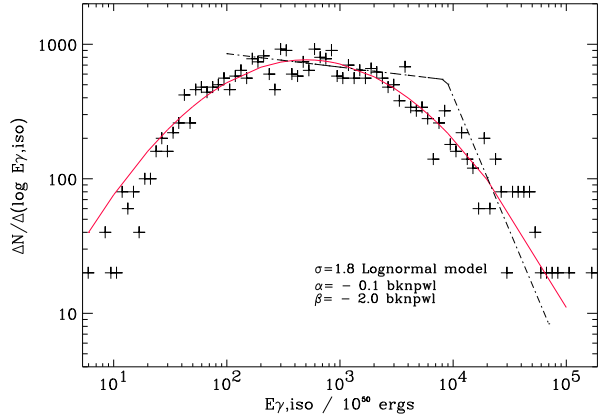,width=8.5truecm}
\caption{`Luminosity' ($E_{\gamma}$) function of BATSE GRB simulated
in this work. Also reported the lognormal fit to the simulated
distribution and the slopes inferred by Guetta et al. (2004) for a
broken power-law representation of the luminosity function.}
\end{figure}

\subsection{Spread of the distributions}

While the above results do support the existence of a connection
between energetics and $E_{\rm peak}$, it is of great relevance both
for understanding the robustness of the physical process behind these
correlations as well as the possible use of GRB for cosmological
studies, to quantitatively determine any intrinsic spread of such
relations.

Indeed Nakar \& Piran (2004) have recently argued that such relations
might be the result of selection effects, as a large number of GRB (at
least 50 per cent of their sample) do not appear to follow the A02
correlation.  Similar findings have been reported by Band \& Preece
(2005) from a more refined analysis, who conclude that 88 per cent of
BATSE bursts are inconsistent with the A02 relation, and only at most
18 per cent qcould follow it.

Whether these results imply that the correlations are totally spurious
- contrary to our indications - or are significantly broader than
estimated so far, has indeed to be determined.

To this aim, we simply allowed for a variable spread ($\sigma$),
approximated as a Gaussian in logarithmic energy, around the A02
correlation.  The comparison of the simulated fluence distributions
with the BATSE ones constrains such Gaussian spread to be centered at
$E_{0}\simeq E_{\rm A02}$ ($\log (E_{0}/E_{\rm A02})=0.05$ for the
bright GRB subsample) and $\sigma =0.17$, the latter fully compatible
with the actual spread in the A02 correlation (see GGL04). While
smaller $\sigma$ are acceptable, a very strong upper limit $\sigma <
0.3$ is imposed by the data: such large spread implies an
excess of GRB both at high and low fluences, arguing against the
possibility that the A02 correlation is in fact just a limit (see
Nakar \& Piran 2004; Band \& Preece 2005).

\begin{figure}
\psfig{file=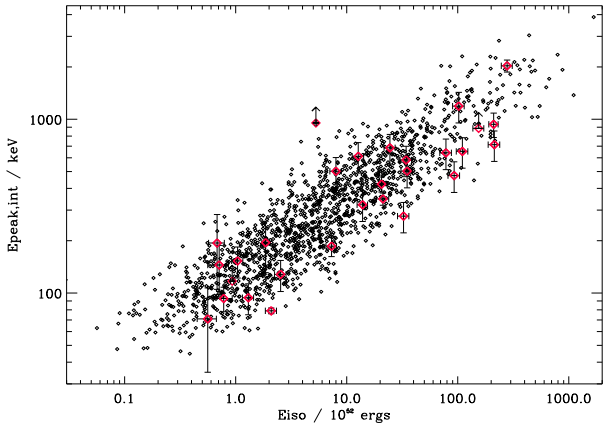,width=8.5truecm}
\caption{Distribution of the simulated GRB in the $E_{\rm peak}$ vs
$E_{\rm iso}$ plane, including the spread around the A02
correlation. The larger symbols indicate the GRB considered by GGL04.}
\end{figure}

\begin{figure}
\psfig{file=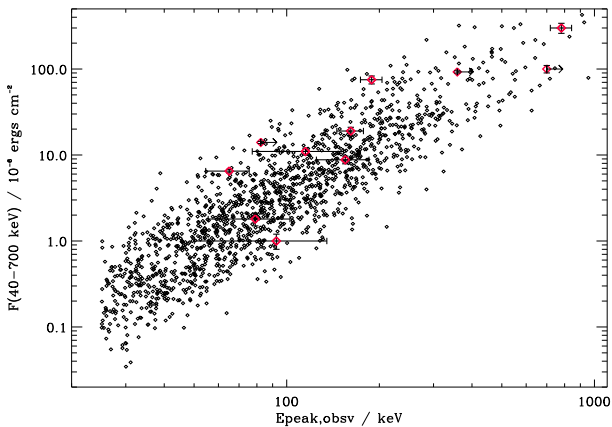,width=8.5truecm}
\caption{Fluence vs $E_{\rm peak}$ distributions as inferred from the
model. The diamonds (larger symbols) indicate the GRB events
considered by GGL04 (for the same energy band), and the black symbols
the GRB reported by Yonetoku et al. (2004).}
\end{figure}

\section{Discussion and Conclusions}

The main result of this work is that the simple assumptions that there
is a link between the energetics and the typical energy of emitted
photons in the prompt phase, as described by the correlations proposed
by A02 and GGL04, and that GRB follow the star formation rate redshift
distribution, are fully consistent with the properties of the `bright'
BATSE GRB (sample by Preece et al. 2000).  For a peak energy
distribution extending to lower frequencies, such a consistency is
found also for the whole of the BATSE population (long GRB). The
$E_{\rm peak}$ distribution, extending to the range of definition of
X-ray rich bursts and X-ray flashes, corresponds to a rising number of
events at lower $E_{\rm peak}$ with respect to that of bright GRB,
flattening and declining below $\sim$ 80 keV.  The extrapolation down
to $E_{\rm peak} \sim$ 200 keV is fully consistent with that sampled
by the GRB analyzed by Yonetoku et al. (2004) at fluences lower that
those of the Preece et al. (2000) GRB.

By adopting an opening angle (lognormal) distribution peaking around
$4-5^o$ and extending to $\sim 8^o$, roughly similar to that observed
(for only $\sim 15$ GRB), consistency with the bright GRB sample is
found also adopting the GGL04 relation. An agreement with the whole
BATSE sample does instead require an opening angle distribution
broader and extending to larger angles (peaking around $\sim 6-8^o$
and up to $\sim 25^o$). Such indication reflects the fact that the A02
and GGL04 distributions have a different slopes, i.e.  the suggestion
of a connection between the GRB opening angle and energetics
$E_{\gamma}$ (and/or $E_{\rm peak}$).  However, our analysis does not
allow to definitely exclude an A02 correlation with slope similar to
the GGL04 one, i.e. an opening angle distribution independent of
energy.

This study indicates that the distribution of fluences of dim GRB
cannot be ascribed to the cosmological distribution of GRB but is
dominated by a spread in $E_{\rm peak}$, and is thus rather
insensitive to the actual GRB redshift distribution at high $z$.  

While these findings of consistency cannot prove the reality of an
intrinsic tight link between the energetics and spectra of GRB, they
significantly corroborate such possibility.  The tested scenario
appears fully consistent. The spread in the above correlations provide
an indication of the strength of the physical connection between the
energetics and spectral properties of the prompt emission, and a
constraint on the statistics required to use GRB as cosmological
distance indicators.

It should be stressed that the above constraints refer only to GRB
observable and observed by the fluence and energy range sensitivity of
BATSE.  Selection effects even within the BATSE sample (related to the
determination of redshift and opening angle) have being indeed claimed
to be responsible for the A02 (and GGL04) correlations by Nakar \&
Piran (2004) and Band \& Preece (2005), on the basis of events not
consistent with them
\footnote{Although it might be difficult to pinpoint a reason why the
GGL04 correlation might be tighter than the A02 one.}.  Two well known
`outliers' of such correlations are provided by two of the GRB with
evidence of an associated Supernova (see also Bosnjak et al. 2005 for
more cases), as well as short GRB (Ghirlanda, Ghisellini \& Celotti
2004). Nakar \& Piran (2004) and Band \& Preece (2005) argue that
actually a large fraction of the whole GRB population does indeed
violate the above relations.

We cannot identify the reason of such discrepancy in the results.
Clearly, it is possible that the agreement we find with the BATSE
fluence distributions is by chance. Alternatively, one could ascribe
it to a significant spread in the above correlations.  However an
estimate of the distribution of the parameter `$d_{\rm k}$' (i.e. the
distance from the A02 correlation, as defined by Nakar \& Piran 2004)
shows that our simulated sample is inconsistent with their findings
within the spread `allowed' by our analysis: the distribution we find
comprises proportionally more GRB with low `$d_{\rm k}$'. In Figure~5
we report the simulated GRB in the $E_{\rm peak}$ vs $E_{\rm iso}$
plane, together with the GRB considered by GGL04 and in Figure~6 the
analogous information in the fluence vs $E_{\rm peak}$ plane.

One aspect worth mentioning, regarding the possibility that the
'outliers' found by the above authors might represent the tail of a
distribution, is the large fraction of high $E_{\rm peak}$ GRB found
by Nakar \& Piran (2004), who estimated $E_{\rm peak} > 250$ keV for
about 50 per cent of their GRB (i.e. corresponding to about 25 per
cent of the whole BASTE long GRB sample). This fraction is not
reproduced in the sample by Yonetoku et al. (2004) whose lower fluence
GRB are typically characterized by softer spectra, supporting our
findings. We stress that our analysis does not suffer from the fluence
(and $z$) limitations required for the estimate of $E_{\rm
peak}$. Unfortunately, the lack of detailed information on the GRB
considered in those two studies does not allow a deeper investigation
on the found discrepancy at this stage.

The direct testing of such correlations based on individual events
requires the determination of redshift (and break time in the
afterglow light curves) for a significant number of GRB.  Indirect
support can however come from the comparison of the inferred $E_{\rm
peak}$ distribution with the extension towards lower energy, to X--ray
rich GRB and X--ray flashes, as will be provided by HETE~2 and Swift.

\section*{Acknowledgments}

We thank Giancarlo Ghirlanda, Gabriele Ghisellini and Claudio Firmani
for showing us the results of their work before submitting their
paper. The Italian MIUR and INAF (ZB and AC) are acknowledged for
financial support.

\end{document}